\documentclass[epsfconf,12pt,eqsecnum,nofootinbib,amsmath,amssymb]{revtex4}
\usepackage{graphicx} 
\usepackage{dcolumn}  
\usepackage{bm}       
\usepackage{latexsym} 
\usepackage{fancyhdr} 

\newcommand{\footnoteskip}{\baselineskip 12pt plus 1pt minus 1pt}
\newcommand{\abstractskip}{\baselineskip 13pt plus 1pt minus 1pt}
\newcommand{\affiliationskip}{\baselineskip 15pt plus 1pt minus 1pt}
\newcommand{\captionskip}{\footnotesize \baselineskip 12pt plus 1pt minus 1pt}

\newcommand{\smGT}{{\scriptscriptstyle >}}
\newcommand{\smLT}{{\scriptscriptstyle <}}

\newcommand{\smI}{{\rm\scriptscriptstyle I}}

\newcommand{\smCoul}{{\rm\scriptscriptstyle coul}}

\begin{document}

\preprint{LA-UR-08-3394}
\pacs{52.20.-j, 52.25.Dg, 52.57.-z}

\title{The energy partitioning of non-thermal particles in a plasma: \\[-8pt]
  or \\[-8pt]
  the Coulomb logarithm revisited }

\author{Robert L. Singleton Jr. and Lowell S. Brown}

\vskip0.2cm 
\affiliation{\affiliationskip
     Los Alamos National Laboratory\\
     Los Alamos, New Mexico 87545, USA
}

\begin{abstract}
\abstractskip
\vskip0.3cm 

\noindent
  The charged particle stopping power in a highly ionized and weakly
  to moderately coupled plasma has been calculated exactly to leading
  and next-to-leading accuracy in the plasma density by Brown,
  Preston, and Singleton (BPS). Since the calculational techniques of
  BPS might be unfamiliar to some, and since the same methodology can
  also be used for other energy transport phenomena, we will review
  the main ideas behind the calculation. BPS used their stopping power
  calculation to derive a Fokker-Planck equation, also accurate to
  leading and next-to-leading orders, and we will also review this.
  We use this Fokker-Planck equation to compute the electron-ion
  energy partitioning of a charged particle traversing a plasma. The
  motivation for this application is ignition for inertial confinement
  fusion --- more energy delivered to the ions means a better chance
  of ignition, and conversely. It is therefore important to calculate
  the fractional energy loss to electrons and ions as accurately as
  possible, as this could have implications for the Laser
  M\'{e}gajoule (LMJ) facility in France and the National Ignition
  Facility (NIF) in the United States.  The traditional method by
  which one calculates the electron-ion energy splitting of a charged
  particle traversing a plasma involves integrating the stopping power
  $dE/dx$. However, as the charged particle slows down and becomes
  thermalized into the background plasma, this method of calculating
  the electron-ion energy splitting breaks down. As a result, the
  method suffers a systematic error of order $T/E_0$, where $T$ is the
  plasma temperature and $E_0$ is the initial energy of the charged
  particle. In the case of DT fusion, for example, this can lead to
  uncertainties as high as 10\% or so. The formalism presented here is
  designed to account for the thermalization process, and in contrast,
  it provides results that are near-exact.
\end{abstract}

\maketitle

\section{Introduction}

When a fast charged particle with initial energy $E_0$ traverses a
plasma, it looses energy at a rate $dE/dx$ per unit of distance, and
it comes into thermal equilibrium after depositing its energy into the
electrons and ions that make up the plasma. Using the formalism of
Brown, Preston, and Singleton (BPS)\,\cite{bps}, which provides a
means of regulating the kinetic equations at short and long distances
in a consistent manner while treating quantum mechanical effects
exactly, we shall calculate this energy splitting in a uniform plasma
to leading and next-to-leading order in the plasma coupling.  Our
motivation is the thermonuclear burn of deuterium and tritium in
inertial confined fusion experiments where a fast $\alpha$ particle is
born with an energy \hbox{$E_0 = 3.54$~MeV}. The more efficiently the
ions are heated, the easier it will be to initiate the bootstrap
heating process that triggers ignition and burn. 

In this paper we shall always work with a plasma whose components have
a single temperature $T$ (measured in energy units). We are currently
generalizing our formalism to a plasma in which the electrons and ions
are at different temperatures\,\cite{bpsEloss}.  The energy
partitioning is usually computed within the context of a fast charged
particle traversing the plasma until coming to a complete stop, in
which case the energy partition into ions and electrons is given by
\begin{equation}
  E_\smI 
  = 
  \int_0^{E_0}\! dE ~ \frac{dE_\smI/dx}{dE/dx} 
\label{eion}
\end{equation}
and  
\begin{equation}
  E_e 
  = 
  \int_0^{E_0}\! dE ~ \frac{dE_e/dx}{dE/dx} \,.
\label{eelectron}  
\end{equation}
Here $dE_\smI/dx$ and $dE_e/dx$ are the stopping power contributions
from the ions and electrons,
\begin{equation}
  \frac{dE}{dx} = \frac{dE_\smI}{dx} + \frac{dE_e}{dx} \,,
\end{equation}
and thus $ E_\smI + E_e = E_0$.  This is only an approximate
description because the fast charged particle does not simply come to
rest within the plasma, but rather it becomes thermalized at the
temperature $T$.  One should not extend the integrals in
Eqs.~(\ref{eion}) and (\ref{eelectron}) down to zero, but rather a
lower limit $E_\text{min}$ on the order of the thermal plasma energy,
$E_\text{min}\sim T$.  Consequently the systematic error in the above
calculation of $E_\smI$ and $E_e$ is of order $T/E_0$, and as we shall
see, the correct electron-ion energy partition relation reads
\begin{equation}
  E_\smI + E_e + \frac{3}{2}\,T= E_0 \ .
\label{EieT0}
\end{equation}

Before examining the energy splitting, we need to discuss the
calculational framework within which it appears. As we shall see, the
correct expression for the energy splitting arises from a
Fokker-Planck equation derived in BPS; however, before jumping into
details, it will be useful to briefly review some salient features of
the stopping power calculation.

\section{The BPS Formalism}

Calculating Coulomb energy exchange processes in a hot plasma is
notoriously difficult because of the subtleties of the Coulomb
interaction, which produce logarithmic divergences at both long and
short distance scales. This problem was first spelled out and solved
to leading order by Landau and then Spitzer in the context of
electron-ion temperature equilibration, and later by Corman {\em et
  al.} for the charged particle stopping power~\cite{landau37,
  spitzer65, corman75}. Since the divergences are only logarithmic,
one introduces {\em ad hoc} short and long distance cutoffs
$b_\text{min}$ and $b_\text{max}$, and the rate of energy loss of some
process (such as temperature equilibration or stopping power) can be
cast in the form
\begin{eqnarray}
  \frac{d{\cal E}}{dt}
  = 
  K \int_{b_\text{min}}^{b_\text{max}} \frac{db}{b}
  = 
  K \hskip-0.35cm 
  \underbrace{\ln\!\left\{\frac{b_{\rm max}}{b_{\rm min}} 
  \right\}}_{\text{Coulomb Logarithm}}  
  \hskip-0.5cm .
\label{Krate}
\end{eqnarray}
The prefactor $K$ is easy to compute exactly. The
logarithmic term, conventionally called the Coulomb logarithm, 
can only be approximated within the above scheme.
The long distance scale $b_\text{max}$ is set by the relevant Debye
screening length, while the short distance scale $b_\text{min}$ is
determined by either the Landau length or the thermal de~Broglie wave
length (or some interpolation between them). As such, this method
suffers a systematic uncertainty in the argument of the Coulomb
logarithm.\footnote{
\footnoteskip
  The constant under the logarithm sometimes varies by an order of
  magnitude from paper to paper within the literature, depending
  upon the choices of $b_\text{min}$ and $b_\text{max}$.
} In the language of perturbation theory, Eq.~(\ref{Krate}) is
accurate to leading order in the plasma coupling constant. This
accuracy was extended to subleading order by BPS\,\cite{bps} which
performed a first principles controlled calculation, including the
exact terms under the logarithm and a rigorous treatment of the
quantum to classical transition.

Reference~\cite{by} discusses at length the manner by which one can
expand thermodynamic quantities such as $d{\cal E}/dt$ as a
perturbation series in powers of a small parameter $g$, the plasma
coupling defined by
\begin{eqnarray}
  g =\frac{e^2 \kappa}{T} \ ,
\label{gdef}
\end{eqnarray}
where $\kappa$ is the Debye wave number. This is just the ratio of the
potential energy of two electrons a Debye distance apart to the
thermal kinetic energy of the plasma, and it is related to the usual
plasma parameter by $g \propto \Gamma^{3/2}$. Quantities expand in
integer powers of $g$, except for possible $\ln g$ terms, and the rate
of energy exchange for the stopping power takes the form
\begin{eqnarray}
  \frac{d{\cal E}}{dt} 
  = 
  -\underbrace{A\, g^2\ln g}_\text{LO}
  \,\,+\, 
  \underbrace{~B\, g^2~}_\text{NLO} \,+\,\,  {\cal O}(g^3) 
  =
  A\, g^2\ln\Lambda_\smCoul \,+\, {\cal O}(g^3) \ .
\label{dedtNLO}
\end{eqnarray}
We have indicated the leading order (LO) and the next-to-leading order
(NLO) terms in the \hbox{$g$-expansion}.  Here $\ln\Lambda_\smCoul =
-\ln\!\left\{ C g\right\}$, with $C$ defined by $B=-A\, \ln C$, and
therefore knowing the next-to-leading order coefficient $B$ is
equivalent to knowing the exact coefficient $C$ under the logarithm.
To get a feel for the numbers, at the center of the sun $g=0.04$, and
the error term in Eq.~(\ref{dedtNLO}) is consequently small.

The problem with directly calculating $A$ and $B$ is that the kinetic
equations diverge and must be regularized in the appropriate
manner. Furthermore, to find the coefficient under the logarithm, this
regularization procedure must preserve the delicate balance between
the long and short distance physics.  Indeed, the BPS calculation~\cite{bps}
includes both short distance physics and {\em dynamic} collective long
distance physics, joined together exactly and unambiguously (and this
is the rub), systematized by a power series expansion in the plasma
coupling constant~$g$. The coefficients are also calculated to all
orders in the dimensionless quantum two-body scattering parameter
$\eta$, thereby providing an exact interpolation between the extrme
classical and quantum regimes.

The rigorous starting point is the BBGKY hierarchy (or its quantum
generalization), which is finite and well defined and does not suffer
from the aforementioned divergences. One must of course truncate this
vast number of equations to something manageable, such as the
Boltzmann or Lenard-Balescu equations, and it is this truncation
process that renders the various three dimensional integrals
divergent. However, as shown in Ref.~\cite{lfirst}, in $\nu$ spatial
dimensions these divergences become simple poles of the form
$1/(\nu-3)$. In spatial dimensions $\nu>3$ the BBGKY hierarchy reduces
to the Boltzmann equation (BE) to leading order in $g$ (the BE is
finite and does not have the usual long distance divergence for
$\nu>3$). Calculating the rate of energy loss using the
$\nu$-dimensional BE gives a result of the form
\begin{eqnarray}
  \frac{d{\cal E}^\smGT}{dt}
  &=& 
  H(\nu)\,\frac{g^2}{\nu-3} 
  +
  {\cal O}(\nu-3) 
  \hskip0.66cm:~  {\rm LO~in}~g~{\rm when~}\nu > 3 \ .
\label{dedtonecal}
\end{eqnarray}
The ``greater-than'' superscript is to remind us that the calculation
has been performed in dimensions $\nu>3$. In a similar manner, to
leading order in $g$ the BBGKY hierarchy reduces to the Lenard-Balescu
equations (LBE) for $\nu<3$ (the LBE is finite and does not suffer
from short distance divergences when $\nu<3$).  A calculation of the
energy rate with the LBE gives a form
\begin{eqnarray}
  \frac{d{\cal E}^\smLT}{dt}
  &=&
  G(\nu)\, \frac{g^{\nu-1}}{3-\nu} 
  + {\cal O}(3-\nu) 
  \hskip0.7cm :~ {\rm LO~in}~g~{\rm when~} \nu < 3 \ .
\label{dedttwocal}
\end{eqnarray}
Note that both rates are of order $g^2$ in three dimensions, and
they both suffer from a divergent simple pole.  The coefficients
$H(\nu)$ and $G(\nu)$ can be expanded in powers of $\epsilon=\nu-3$,
with
\begin{eqnarray}
  H(\nu) 
  =
  -A + \epsilon \,H_1 + {\cal O}(\epsilon^2)
  ~~~{\rm and}~~~
  G(\nu) 
  =
  -A + \epsilon \,G_1 + {\cal O}(\epsilon^2) \ .
\label{Gexp}
\end{eqnarray}
The leading terms must be equal, $H(3)=G(3)=-A$. This arises from the
calculation itself and is not imposed by hand, and it makes the short-
and long-distance poles cancel, thereby giving a finite result.

Since the rates $d{\cal E}^\smGT/dt$ and $d{\cal E}^\smLT/dt$ were
calculated in mutually exclusive dimensional regimes, one might think
that they cannot be compared. However (and this is perhaps the most
crucial step in the method, and certainly the most subtle), we can
analytically continue the quantity $d{\cal E}^\smLT/dx$ to dimensional
values $\nu>3$, after which we can directly compare the rates
(\ref{dedtonecal}) and (\ref{dedttwocal}) in a common dimension $\nu$,
and the limit $\nu \to 3$ may then be taken. Upon writing the
$g$-dependence of Eq.~(\ref{dedttwocal}) as $g^{2 + (\nu-3)}$, when
$\nu>3$ we see that the rate (\ref{dedttwocal}) is indeed higher order
in $g$ than Eq.~(\ref{dedtonecal}) since $\epsilon=\nu-3 > 0$:
\begin{eqnarray}
  \frac{d{\cal E}^\smLT}{dt}
  &=&
  -G(\nu)\, \frac{g^{2+\epsilon}}{\nu-3} 
  +
  {\cal O}(\nu-3) 
  ~~:~{\rm NLO~in}~g~{\rm when~} \nu > 3 \ .
\label{NLOgterm}
\end{eqnarray}
The individual pole-terms in Eqs.~(\ref{dedtonecal}) and
(\ref{NLOgterm}) will cancel giving a finite result when the leading
and next-to-leading order terms are added. Summing terms
(\ref{dedtonecal}) and (\ref{NLOgterm}), using the relation
$g^{\epsilon} = \exp\{\epsilon \ln g\} = 1 + \epsilon\ln g + {\cal
O}(\epsilon^2)$, and taking the $\epsilon \to 0$ limit gives
\begin{eqnarray}
  \frac{d{\cal E}}{dt}
  =
  - A\, g^2 \ln g   +  B\, g^2  + {\cal O}(g^3) \ , 
\end{eqnarray}
with $B=H_1-G_1$. This is in agreement with Eq.~(\ref{dedtNLO}).  In
this way, BPS has calculated the charged particle stopping power
accurate to leading order and next-to-leading order in $g$.

BPS also derived a Fokker-Planck equation accurate to leading and
next-to-leading order in the plasma coupling\,\cite{bps}.  Denoting
the phase space density for the dilute collection of charged particles
by $f({\bf r}, {\bf p}, t)$, this equation reads
\begin{equation}
  \left[ \frac{\partial}{\partial t} + {\bf v} \cdot \nabla \right]
   f({\bf r},{\bf p},t)  
  = 
  {\sum}_b\,
  \frac{\partial}{\partial p^k} \, C_b^{k\ell}({\bf p}) 
  \left[\beta\, v^\ell + \frac{\partial}{\partial p^\ell} \right] 
  f({\bf r},{\bf p},t) \,,
\label{FP}
\end{equation}
where the sum runs over the plasma components $b$, $\beta=1/T$, the
vector ${\bf v} = {\bf p} / m $ is the velocity of a particle with
momentum ${\bf p}$, and the summation convention is used for repeated
vector indices.  The symmetric tensor $C_b^{k\ell}$ has longitudinal
and transverse components,
\begin{equation}
  C_b^{k\ell}({\bf p}) 
  = 
  {\cal A}_b(E) \, \frac{\hat v^k \hat v^\ell}{\beta v} +
  {\cal B}_b(E) \, \frac{1}{2} \, \left( \delta^{k \ell} - 
  \hat v^k \hat v^\ell \right) \,,
\label{decomp}
\end{equation}
where $v = | {\bf v}|$ and $ \hat {\bf v} = {\bf v} / v $, and
$E=\frac{1}{2}\,m v^2$. 
  We denote
the sum of the ion components of the ${\cal A}$-coefficients by ${\cal
  A}_\smI={\sum}_i{\cal A}_i$ and the electron component by ${\cal
  A}_e$, with ${\cal A} = {\cal A}_\smI + {\cal A}_e$. Expressions for
the ${\cal A}_b$ can be found in BPS\,\cite{bps}.  With our
conventions, the number and kinetic energy densities of the charged
particles is given by
\begin{eqnarray}
  n({\bf r}, t) 
  &=& 
  \int\frac{d^3{\bf p}}{(2\pi\hbar)^3} \,  f({\bf r},{\bf p},t) 
\label{number}
\end{eqnarray}
and
\begin{eqnarray}
  {\cal E}({\bf r}, t) 
  &=& 
  \int \frac{d^3 {\bf p}}{(2\pi\hbar)^3} \,
  \frac{{\bf p}^2}{2m} \, f({\bf r},{\bf p},t) \ .
\label{FPenergy}
\end{eqnarray}

We can derive a relation between the stopping power and the ${\cal
  A}$-coefficients  in the following manner. For a single particle at
${\bf r}_p$ moving with velocity ${\bf v}_p$, the distribution
function takes the form $f({\bf r},{\bf p},t) = (2\pi\hbar)^3
\delta({\bf r} - {\bf r}_p) \delta({\bf p} - {\bf p}_p)$, and the
Fokker-Planck equation gives the particle's rate of energy loss as
\begin{equation}
  v_p \,\frac{dE}{dx}
  =
  \frac{dE}{dt} 
  =   
  {\sum}_b
  \left[ \beta v^\ell_p - \frac{\partial}{\partial p^\ell_p} \right] \, 
  v^k_p \, C_b^{k\ell}({\bf p}_p) \,.
\label{dedtckl}
\end{equation}
Upon substituting the decomposition (\ref{decomp}) for the scattering
tensor and dropping the projectile subscript $p$, the contribution
from species $b$ appears as
\begin{equation}
  \frac{dE_b}{dx} =
  \left[1 - \frac{1}{\beta m v} \, 
  \frac{\partial}{\partial v^\ell} \, \hat v^\ell \right]
  {\cal A}_b(E) \ .
\label{dedtb} 
\end{equation}
As $v$ gets large ($\beta m v^2 \gg 1$) note that $dE_b/dx \to {\cal
  A}_b$.

\section{Formulation of the Problem}

\begin{figure}[b]
\includegraphics[scale=0.5]{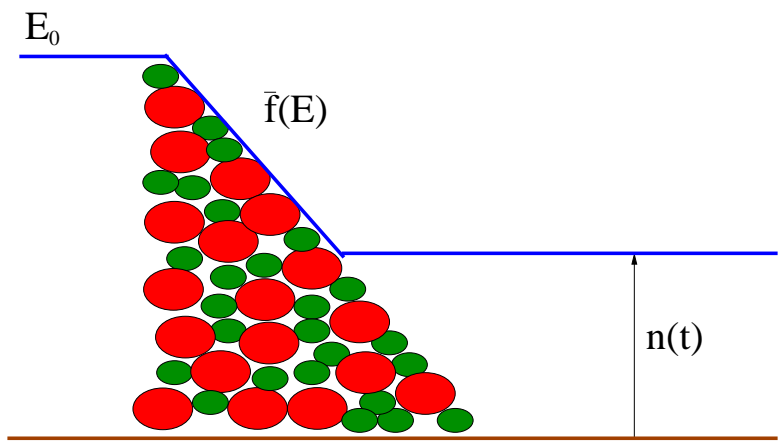}
\caption{\footnoteskip  
  The waterfall analogy. The small green rocks represent the plasma
  electrons while the larger red rocks are the plasma ions, with the
  flowing `water' in blue representing the evolution of the produced
  charged particles ($\alpha$ particles for DT fusion). As `water'
  falls down the electron-ion slope at a constant rate determined by
  $\bar f(E)$, energy is deposited into electrons and ions. At the
  bottom of the fall is a lake into which the excess `water' drains,
  representing the final thermalized particles, with height $n(t)$
  and Maxwell-Bolztmann distribution $\propto e^{-\beta E}$. 
}
\label{fig:name}
\end{figure}

With this background in hand, we turn now to the energy splitting
problem. Rather than tracking an individual charged particle slowing
down in the plasma, it is much simpler to examine a homogeneous and
isotropic source of charged particles of a single energy $E_0$.  The
distribution function will therefore depend only upon the energy and
time, $f=f(E,t)$. Furthermore, the homogeneity and
isotropy conditions will greatly simplify the form of the
Fokker-Planck equation: only the transverse coefficients ${\cal A}_b$
will enter the diffusion kernel on the right-hand side of Eq.~(\ref{FP}),
while the convective term on the left-hand side will of course vanish.
Thus, we now consider the inhomogeneous Fokker-Planck equation with a
source, which, after some algebra reads
\begin{equation}
  \left\{\frac{\partial}{\partial t} 
  - 
  \frac{2}{m v}\, \frac{\partial}{\partial E}\,
  {\cal A}(E)\, E \left[ 1 + 
  T\,\frac{\partial}{\partial E}
  \right] \right\} \, f(E,t) 
  = \delta\left( E - E_0 \right) \, s(t)  \ .
\label{inhomo}
\end{equation}
We find\,\cite{bpsEloss} that the relevant solution is of the form
\begin{equation}
  f(E,t) = n(t)\,{\cal N}\, e^{-\beta E}  + \bar f(E) \ .
\label{asymp}
\end{equation}
We proceed now to motivate the structure of the solution (\ref{asymp})
and the nature of the time independent function ${\bar f}(E)$.  The
particles will eventually thermalize to a Maxwell-Boltzmann
distribution, and the first term of Eq.~(\ref{asymp}) merely
represents the buildup and subsequent thermalization of the particles
produced by the source.  The normalization factor ${\cal N}$ is chosen
so that $n(t)$ is the number density of the produced particles once
they have thermalized into the Maxwell-Boltzmann distribution $\propto
e^{-\beta E}$.  The time-independent piece $\bar f(E) $ describes the
steady state of nonthermal particles losing energy to the plasma, {\em
  i.e.} particles cascading down ``energy bins'' from the initial
energy $E_0$ to the final thermal energy.  The situation described
here can be pictured as the flow of water over a rocky waterfall that
slows the motion of the water as it descends.  The initial rate of
flow of the river corresponds to the rate of produces particles; the
height of the waterfall to the initial energy $E_0$. The energy
dissipated in the fall corresponds to the energy lost to the ions and
electrons and is determined by $\bar f(E)$.  The final flow into a
horizontal lake corresponds to the build up of the particles into
their final thermal equilibrium state. This is illustrated in
Fig.~\ref{fig:name}.

We now specialize to the relevant case in which the source $s(t)$
slowly turns on and attains a constant value. In this case, particles
become produced at a constant rate per unit volume $\dot n_0$, and
from Eq.~(\ref{FPenergy}) the rate of change in energy density becomes
\begin{equation}
  \dot {\cal E}(t) 
  = 
  E_0 \, \dot n_0 - \int\frac{d^3{\bf p}}{(2\pi\hbar)^3}\, 
  \left[ {\cal A}_\smI + {\cal A}_e \right]\,  
  \left[ v + \frac{1}{\beta m }\,\frac{\partial}{\partial v} 
  \right]\, \bar f(E) \,.
\label{Edot}
\end{equation}
The fraction of energy lost to the ions and electrons is now
identified as\,\cite{bpsEloss}
\begin{eqnarray}
  \frac{E_\smI}{E_0} 
  &=& 
  \frac{1}{\dot n_0 \, E_0}\int\frac{d^3{\bf p}}{(2\pi\hbar)^3}\,  
  {\cal A}_\smI \,  \left[ v + \frac{1}{\beta m }\,
  \frac{\partial}{\partial v} \right]  \, \bar f(E) 
\label{Ifrac}
\end{eqnarray}
and
\begin{eqnarray}
  \frac{E_e}{E_0} 
  &=& 
  \frac{1}{\dot n_0 \, E_0} \int\frac{d^3{\bf p}}{(2\pi\hbar)^3}\,  
  {\cal A}_e\,  \left[ v + \frac{1}{\beta m }\,
  \frac{\partial}{\partial v} \right]  \, \bar f(E) \ .
\label{efrac}
\end{eqnarray}
In the steady state, the energy density build up of final particles is
their thermal energy per particle times the increase in number
density, $\dot {\cal E} = \frac{3}{2} \, T \, \dot n_0$.  And so, in
view of Eqs.~(\ref{Ifrac}) and (\ref{efrac}), the energy balance expression
(\ref{Edot}) now appears as
\begin{equation}
  \frac{3}{2} \, T \, \dot n_0 
  = 
  \left[ E_0 - E_\smI - E_e \right] \, \dot n_0 \,,
\label{obvious}
\end{equation}
which gives expression (\ref{EieT0}): the original energy of a
produced particle is lost to the ions and electrons with the remainder
being the thermal energy of a free particle.

We can simplify Eqs.~(\ref{Ifrac}) and (\ref{efrac}) by calculating the
action of the operator in square brackets on $\bar f(E)$, and one
finds\,\cite{bpsEloss}
\begin{eqnarray}
\frac{E_\smI}{E_0} &=& 
\int_0^{E_0} \frac{dE}{E_0} \, \frac{ {\cal A}_\smI(E)}{{\cal A}(E) }
\left[ {\rm erf}(\sqrt{\beta E} ) -  
\sqrt{ \frac{ 4\beta  E }{\pi} }  \,  e^{-\beta E} \right]
\label{IIIanswer}
\end{eqnarray}
and
\begin{eqnarray}
\frac{E_e}{E_0} &=& 
\int_0^{E_0} \frac{dE}{E_0} \, \frac{ {\cal A}_e(E)}{{\cal A}(E) }
\left[ {\rm erf}(\sqrt{\beta E} ) -  
\sqrt{ \frac{ 4\beta  E }{\pi} }  \,  e^{-\beta E} \right] \ ,
\label{eeeanswer}
\end{eqnarray}
where ${\rm erf}(x)$ is the error function. Since $dE_b/dx \to {\cal
  A}_b$ for large energies, and the error function approaches unity,
at high energies $E$ we see that Eqs.~(\ref{IIIanswer}) and
(\ref{eeeanswer}) approach the same form as the more intuitive but
less accurate results (\ref{eion}) and (\ref{eelectron}). The
primary differences occur for $E \sim T$.

\begin{figure}[b]
\includegraphics[scale=0.45]{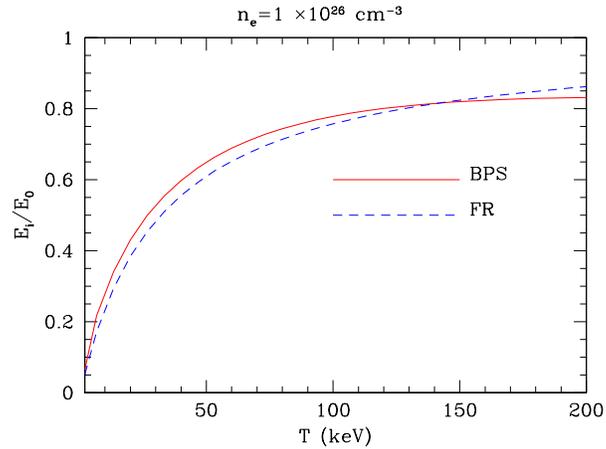}
\vskip-1cm 
\caption{\captionskip
  The fractional energy loss into ions as a function of the plasma
  temperature for an $\alpha$ particle in an equimolar DT plasma with
  initial energy $E_0=3.54\,{\rm MeV}$. The electrons and ions have a
  common temperature $T$ and the electron number density of the plasma
  is $n_e=1 \times 10^{26}\,{\rm cm}^{-3}$.  The solid line is the
  analytic calculation of this work, and the dashed line is the fit
  provided by Fraley {\em et al.}}
\label{fig:frac26}
\end{figure}

\begin{figure}[!]
\includegraphics[scale=0.45]{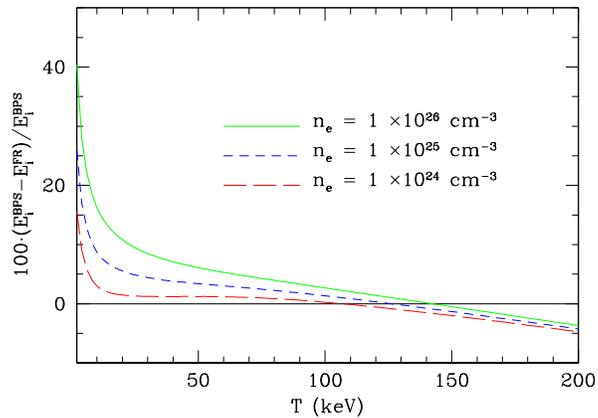}
\vskip-1cm 
\caption{\captionskip
  The percent change of the ion fractional energy loss between Fraley
  {\em et al.} and BPS for a 3.54\,MeV $\alpha$ particle in an
  equimilar DT plasma for three densities $n_e=10^{24} \,,\,
  10^{25}\,,\,10^{26}\,{\rm cm}^{-3}$.  }
\label{fig:change}
\end{figure}

Let us compare these near-exact results with the less precise but well
known result of Fraley {\em et al.}\,(FR)\,\cite{fraley}.  Starting
with a phenomenological model of the stopping power, these authors
show that the simple rule
\begin{equation}
  \frac{E_\smI}{E_0} = \frac{1}{1 + 32 \, {\rm keV} \, / T_e} 
\end{equation}
provided a good fit to their results. Figure~\ref{fig:frac26} shows
the fraction energy loss to ions for BPS and FR for a DT plasma with
electron number density $n_e=1 \times 10^{26}\,{\rm cm}^{-3}$. We see
that FR somewhat underestimates the energy deposited to ions for
temperatures up to around 100\,{\rm keV}, and FR slightly over
estimates $E_\smI$ for larger energies.  In Fig.~\ref{fig:change} we
compare the percent difference between BPS and FR over a wide range of
densities.

\vskip-1cm 

\vskip1cm 



\begin{thebibliography}{99}
\baselineskip16pt plus 1pt minus 1pt

\bibitem{bps}
  L.~S.~Brown, D.~L.~Preston, and R.~L.~Singleton~Jr.,
  Phys. Rep. {\bf 410} (2005) 237-333, arXiv:physics/0501084.

\bibitem{bpsEloss}
    L.~S.~Brown, D.~L.~Preston, and R.~L.~Singleton~Jr.,
    LA-UR-07-8096. 

\bibitem{landau37}
  L. D. Landau,  Phys. Z. Sowjetunion 10, (1936) 154; 
  Sov. Phys. JETP 7 (1937) 203. 

\bibitem{spitzer65}
  L. Spitzer,  The Physics of Fully Ionized Gasses, Interscience
  Publishing (New York) 1965.

\bibitem{corman75}
  E.G. Corman, W.E. Loewe, G.E. Cooper, and A.M. Winslow, Nuclear 
  Fusion 15 (1975) 377.

\bibitem{by} 
  L.~S. Brown and L.~G. Yaffe, 
  Phys. Rep. {\bf 340} (2001) 1-164, 
  physics/9911055. 

\bibitem{lfirst}
  L. S. Brown, 
  Phys. Rev. {\bf D 62} (2000) 045026,
  arXiv:physics/9911056.

\bibitem{bpsexplained}
  R.L.~Singleton~Jr., {\em BPS Explained I: Temperature Relaxation in a
  Plasma}, arXiv: 0706.2680;  
  {\em BPS Explained II: Calculating the Equilibration Rate
  in the Extreme Quantum Limit}, arXiv: 0712.0639.

\bibitem{fri} 
  E. A. Frieman and D. L. Book, 
  Phys. Fluids {\bf 6} (1963) 1700.

\bibitem{wei} 
  J. Weinstock, 
  Phys. Rev.  {\bf 133} (1966) A673.

\bibitem{gould} 
  H. A. Gould and H. E. DeWitt, 
  Phys. Rev. {\bf 155} (1966) 68.

\bibitem{fraley}
  G.S.~Fraley, E.J.~Linnebur, R.J.~Mason, and R.L.~Morse,
  Phys. Fluids {\bf 17} (1974) 474.



\end{thebibliography}
\end{document}